\documentclass[12pt,english]{article}
\usepackage[T1]{fontenc}
\usepackage[latin9]{inputenc}
\usepackage{geometry}
\usepackage{fancyhdr}
\usepackage{babel}
\usepackage{textcomp}
\usepackage{url}
\usepackage{amsmath}
\usepackage{amsthm}
\usepackage{amssymb}
\usepackage{graphicx}
\usepackage{setspace}
\usepackage[authoryear]{natbib}
\PassOptionsToPackage{normalem}{ulem}
\usepackage{ulem}
\usepackage[unicode=true]
 {hyperref}

\makeatletter
\theoremstyle{definition}
\newtheorem{defn}{\protect\definitionname}
\theoremstyle{plain}
\newtheorem{thm}{\protect\theoremname}
\theoremstyle{plain}
\newtheorem{lyxalgorithm}{\protect\algorithmname}

\usepackage{setspace}
\usepackage{pdflscape}
\usepackage{graphicx}
\usepackage{url}

\makeatother

\providecommand{\algorithmname}{Algorithm}
\providecommand{\definitionname}{Definition}
\providecommand{\theoremname}{Theorem}

\begin{document}

\lhead{UC - Berkeley and St. Gallen}

\rhead{Bryan S. Graham and Andrin Pelican}

\rfoot{© B. Graham and A. Pelican, 2019}

\begin{singlespacing}
\title{Testing for Externalities in Network Formation Using Simulation}

\maketitle
\medskip{}

\begin{center}
{\large{}Bryan S. Graham}\footnote{{\footnotesize{}Department of Economics, University of California
- Berkeley, 530 Evans Hall \#3380, Berkeley, CA 94720-3880 and National
Bureau of Economic Research, }{\footnotesize{}\uline{e-mail:}}{\footnotesize{}
\href{http://bgraham\%40econ.berkeley.edu}{bgraham@econ.berkeley.edu},
}{\footnotesize{}\uline{web:}}{\footnotesize{} \url{http://bryangraham.github.io/econometrics/}.
}\\
{\large{}$^{\dagger}$}{\footnotesize{}Department of Economics, School
of Economics and Political Science, University of St. Gallen, }{\footnotesize{}\uline{e-mail:}}{\footnotesize{}
\href{http://andrin.pelican\%40student.unisg.ch}{andrin.pelican@student.unisg.ch}.\smallskip{}
}\\
{\footnotesize{}We thank Aureo de Paula and participants in the Mannheim/Bonn
summer school on social networks for useful feedback. All the usual
disclaimers apply. Financial support from NSF grant SES \#1851647
is gratefully acknowledged.}}{\large{}and Andrin Pelican$^{\dagger}$}{\large\par}
\par\end{center}

\begin{center}
\medskip{}
\textsc{\large{}\uline{Initial Draft}}\textsc{\large{}: June 2019,
}\textsc{\large{}\uline{This Draft}}\textsc{\large{}: July 2019}{\large\par}
\par\end{center}

\begin{center}
{\Large{}Abstract}{\Large\par}
\par\end{center}

{\footnotesize{}\smallskip{}
}{\footnotesize\par}

We discuss a simplified version of the testing problem considered
by \citet{Pelican_Graham_WP2019}: testing for interdependencies in
preferences over links among $N$ (possibly heterogeneous) agents
in a network. We describe an exact test which conditions on a sufficient
statistic for the nuisance parameter characterizing any agent-level
heterogeneity. Employing an algorithm due to \citet{Blitzstein_Diaconis_IM11},
we show how to simulate the null distribution of the test statistic
in order to estimate critical values and/or p-values. We illustrate
our methods using the Nyakatoke risk-sharing network. We find that
the transitivity of the Nyakatoke network far exceeds what can be
explained by degree heterogeneity across households alone.

\qquad{}\textbf{JEL Classification:}

{\footnotesize{}\smallskip{}
}{\footnotesize\par}

\qquad{}\textbf{Keywords:} Networks, Strategic Interaction, Testing,
$\beta$ Model, Importance Sampling

\end{singlespacing}

\thispagestyle{empty} \pagebreak{}

\citet{deWeerdt_IAP04} studies the formation of risk-sharing links
among $N=119$ households located in the Tanzanian village of Nyakatoke.
Let $\hat{P}\left(\vcenter{\hbox{\includegraphics[scale=0.125]{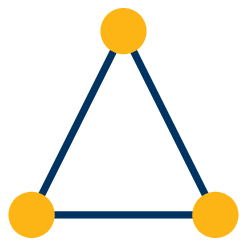}}}\right)$
and $\hat{P}\left(\vcenter{\hbox{\includegraphics[scale=0.125]{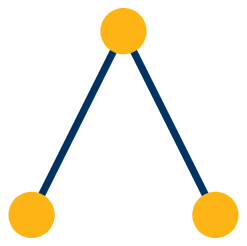}}}\right)$,
respectively, equal the fraction of all $\tbinom{N}{3}$ triads which
take the triangle and two-star configuration. \citet{deWeerdt_IAP04}
finds that the transitivity of the Nyakatoke risk-sharing network
\begin{equation}
\mathrm{\hat{TI}}=\frac{3\hat{P}\left(\vcenter{\hbox{\includegraphics[scale=0.125]{triangle.eps}}}\right)}{3\hat{P}\left(\vcenter{\hbox{\includegraphics[scale=0.125]{triangle.eps}}}\right)+\hat{P}\left(\vcenter{\hbox{\includegraphics[scale=0.125]{twostar.eps}}}\right)}=0.1884\label{eq: transitivity_index}
\end{equation}
is almost three times its density ($\hat{P}\left(\vcenter{\hbox{\includegraphics[scale=0.125]{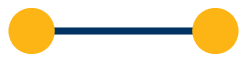}}}\right)=0.0698$).
There are several, not necessarily mutually exclusive, explanations
for the clustering present in Nyakatoke. First, it may reflect the
increased value of a risk-sharing relationship between two households
when ``supported'' by a third household \citep{Jackson_et_al_AER12}.
Households $i$ and $j$ may value a risk sharing link with each other
more if they are both additionally linked to household $k$. Household
$k$ may then serve as a monitor, arbiter and referee for households
$i$ and $j$'s relationship (increasing its value). Second, clustering
may simply be an artifact of degree heterogeneity, whereby high degree
households link more frequently with one-another and hence form triangles
($\vcenter{\hbox{\includegraphics[scale=0.125]{triangle.eps}}}$)
incidentally. Third, clustering may also stem from homophilous sorting
on kinship, ethnicity, religion and so on \citep{McPherson_et_al_ARS01}.

In this chapter we outline a method of randomization inference, in
the spirit of \citet{Fisher_DesignBook35}, that can be used to ``test''
for externalities, or interdependencies, in link formation.\footnote{We place ``test'' in quotes to emphasize that while, as in other
examples of randomization inference, the null hypothesis is explicitly
formulated, rejection may arise for a variety of reasons. See \citet[Chapter 3]{Cox_Bk06}
for additional examples and discussion.} Externalities arise when the utility two agents generate when forming
a link varies with the presence or absence of edges elsewhere in the
network. Attaching greater utility to ``supported'' links is one
such interdependency. Externalities typically imply that an efficient
network will not form when link formation is negotiated bilaterally
\citep[e.g.,][]{Bloch_Jackson_JET07}. Consider household $i$, who
is linked to $k$, an incidental consequence of $i$ additionally
linking to $j$, is that she will then be able to ``support'' any
link between $j$ and $k$. When evaluating whether to form a link
with $j$, $i$ may not internalize this additional, non-bilateral,
benefit; consequently, there may be too few risk-sharing links in
Nyakatoke.

The scientific and policy implications of externalities are considerable
\citep[cf.,][]{Graham_NBER16}. In the presence of interdependencies
small re-wirings of a network may induce a cascade of link revisions,
alternating network structure, as well as agents' utility, substantially.
In the absence of interdependencies links form bilaterally and re-wirings
will not induce additional changes in network structure.

\begin{figure}
\begin{centering}
\caption{\label{fig: Nyakatoke-risk-sharing-network}Nyakatoke risk-sharing
network}
\includegraphics{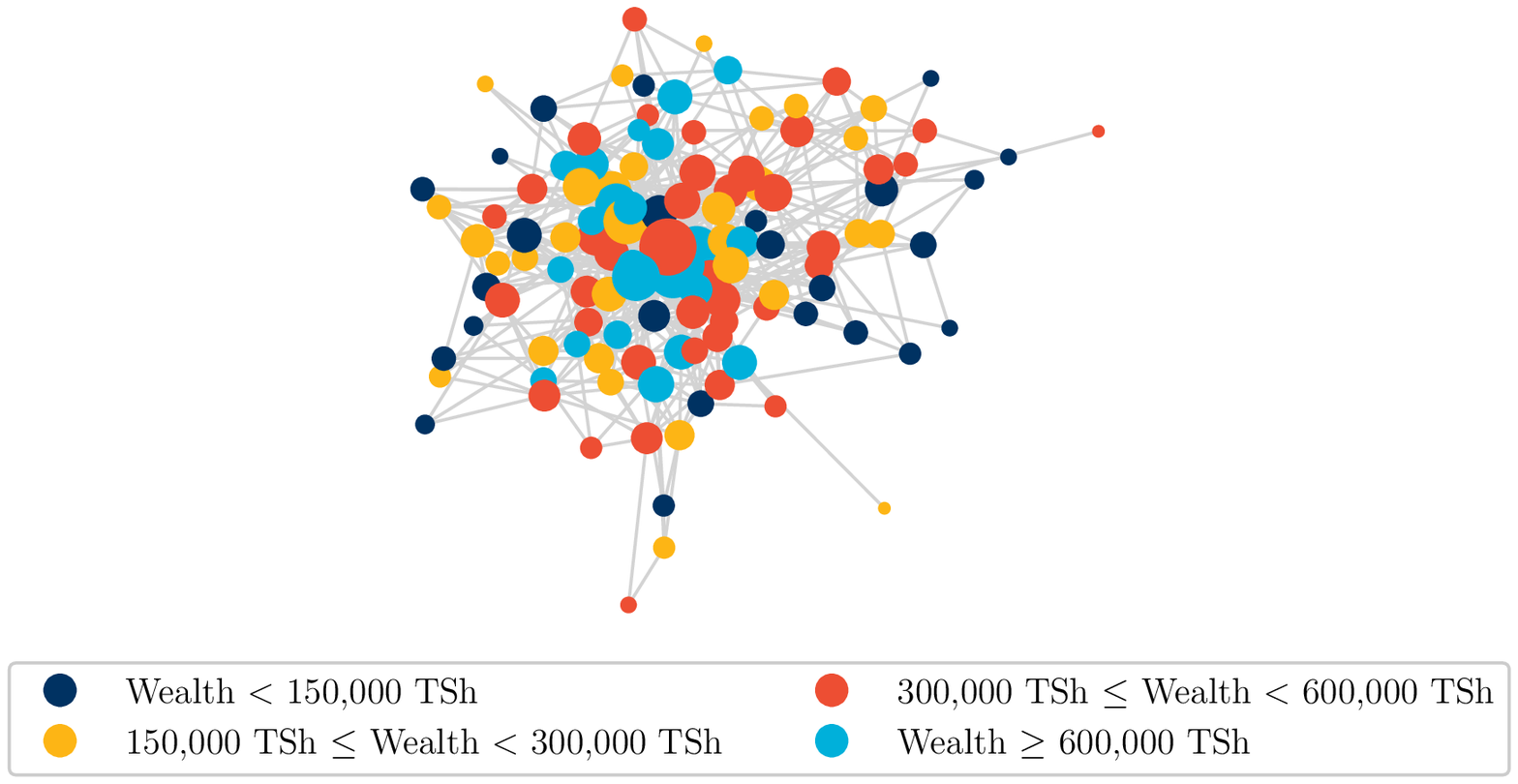}
\par\end{centering}
\textsc{\small{}\uline{Notes:}}{\small{} Nodes/Households are colored
according to total land and livestock wealth in Tanzanian Shillings
(calculated as described by \citet{Comola_Fafchamps_EJ14}). Node
sizes are proportional to degree.}{\small\par}

\textsc{\small{}\uline{Source:}}{\small{} \citet{deWeerdt_IAP04}
and authors' calculations. Raw data available at \url{https://www.uantwerpen.be/en/staff/joachim-deweerdt/}
(Accessed January 2017).}{\small\par}
\end{figure}

Testing for externalities in network formation is not straightforward;
particularly if the null model allows for rich forms of agent-level
heterogeneity. Allowing for heterogeneity under the null is essential,
after all, such heterogeneity provides an alternative explanation
for the type of link clustering which arises in the presence of externalities.

The approach outlined here draws upon \citet{Pelican_Graham_WP2019}.
Our goal here, however, is primarily pedagogical; to introduce some
key ideas and methods likely to be of interest to empirical researchers
interested in networks. Relative to what is presented here, \citet{Pelican_Graham_WP2019}
additionally consider (i) directed as well as undirected networks,
(ii) null models which allow for both degree heterogeneity \emph{and}
homophily, (iii) models with \emph{and} without transfers, (iv) constructing
test statistics to maximize power in certain directions and (v) novel
Markov Chain Monte Carlo (MCMC) methods of simulating the null distribution
of the test statistic.

While the presentation in this chapter is more basic, the upshot is
that the method outlined here can be implemented by combining familiar
heuristics \citep[e.g.,][]{Milo_el_al_Sci02} with a little game theory
and the adjacency matrix simulation algorithm of \citet{Blitzstein_Diaconis_IM11}.

We begin by outlining a simple network formation game with transfers
in the spirit of \citet{Bloch_Jackson_JET07}. We then formally describe
our test. Next we discuss how to simulate the null distribution of
our test statistic using an algorithm due to \citet{Blitzstein_Diaconis_IM11};
we also briefly discuss other applications of this algorithm as well
as alternatives to it. We close with an illustration based upon the
Nyakatoke network.

In what follows we let $\mathbf{D}$ be the adjacency matrix associated
with a simple graph $G\left(\mathcal{V},\mathcal{E}\right)$ defined
on $N=\left|\mathcal{V}\right|$ vertices or agents. The set $\mathcal{V}=\left\{ 1,\ldots,N\right\} $
includes all agents in the network and $\mathcal{E}\subseteq\mathcal{V}\times\mathcal{V}$
equals the set of undirected edges or links among them. The set of
all $N\times N$ undirected adjacency matrices is denoted by $\mathbb{D}_{N}.$

\section*{A strategic network formation game with transfers}

In this section we outline a simple model of strategic network formation
where agents may make (bilateral) transfers to one another. Let $\nu_{i}\thinspace:\thinspace\mathbb{D}_{N}\rightarrow\mathbb{R}$
be a utility function for agent $i$, which maps adjacency matrices
-- equivalently networks -- into utils. The marginal utility for
agent $i$ associated with (possible) edge $\left(i,j\right)$ is
\begin{equation}
MU_{ij}\left(\mathbf{D}\right)=\left\{ \begin{array}{cc}
\nu_{i}\left(\mathbf{D}\right)-\nu_{i}\left(\mathbf{D}-ij\right) & \text{if}\thinspace D_{ij}=1\\
\nu_{i}\left(\mathbf{D}+ij\right)-\nu_{i}\left(\mathbf{D}\right) & \text{if}\thinspace D_{ij}=0
\end{array}\right.\label{eq: marginal_utility}
\end{equation}
where $\mathbf{D}-ij$ is the adjacency matrix associated with the
network obtained after deleting edge $\left(i,j\right)$ and $\mathbf{D}+ij$
the one obtained via link addition.

From \citet{Bloch_Jackson_IJGT06}, a network is \emph{pairwise stable
with transfers} if the following condition holds.
\begin{defn}
\label{def: Pairwise-stability}(\textsc{Pairwise stability with Transfers)
}The network $G\left(\mathcal{V},\mathcal{E}\right)$ is pairwise
stable with transfers if \\
(i) $\forall\left(i,j\right)\in\mathcal{E}\left(G\right),\thinspace MU_{ij}\left(\mathbf{D}\right)+MU_{ji}\left(\mathbf{D}\right)\geq0$\\
(ii) $\forall\left(i,j\right)\notin\mathcal{E}\left(G\right),\thinspace MU_{ij}\left(\mathbf{D}\right)+MU_{ji}\left(\mathbf{D}\right)<0$
\end{defn}
Definition \ref{def: Pairwise-stability} states that the marginal
utility of all links actually present in a pairwise stable network
is (weakly) positive, while that associated with links not present
is negative. The definition presumes utility is transferable, since
it only requires that the \emph{sum} of utility to $i$ \emph{and}
$j$ associated with edge $\left(i,j\right)$ is positive. When the
sum of the two marginal utilities is positive, there exists a within-dyad
transfer such that both agents benefit. Note also that the definition
is pairwise: the benefit of a link is evaluated conditional on the
remaining network structure. It is possible, for example, that a coalition
of players could increase their utility by jointly forming a set of
links and making transfers to one another. Imposing more sophisticated
play of this type would result in a refinement relative to the set
of network configurations which satisfy Definition \ref{def: Pairwise-stability}
\citep[cf.,][]{Jackson_NetBook08}.

In this chapter we will specialize to utility functions of the form
\begin{equation}
\nu_{i}\left(\mathbf{d},\mathbf{U};\gamma_{0},\delta_{0}\right)=\sum_{j}d_{ij}\left[A_{i}+B_{j}+\gamma_{0}s_{ij}\left(\mathbf{d}\right)-U_{ij}\right]\label{eq: utility_function}
\end{equation}
with $\mathbf{U}=\left[U_{ij}\right]$. Here $A_{i}$ and $B_{j}$
capture agent-level degree heterogeneity \citep[cf.,][]{Graham_EM17}.
If $A_{i}$ is high, then the baseline utility associated with any
link is high for agent $i$ ($i$ is an ``extrovert''). If $B_{j}$
is high, then $j$ is a particularly attractive partner for all other
agents ($j$ is ``popular''). We leave the joint distribution of
$\mathbf{A}=\left[A_{i}\right]$ and $\mathbf{B}=\left[B_{i}\right]$
unrestricted in what follows.

The $s_{ij}\left(\mathbf{d}\right)$ term is associated with externalities
in link formation. We require that $s_{ij}\left(\mathbf{d}\right)=s_{ij}\left(\mathbf{d}-ij\right)=s_{ij}\left(\mathbf{d}+ij\right)$;
additional restrictions might be needed to ensure the existence of
a network that is pairwise stable with transfers and/or a test statistic
with a non-degenerate null distribution.

Instead of formulating additional high-level conditions on $s_{ij}\left(\mathbf{d}\right)$,
in what follows we emphasize, and develop results for, two specific
examples. In the first $s_{ij}\left(\mathbf{d}\right)$ equals
\begin{equation}
s_{ij}\left(\mathbf{d}\right)=\sum_{k}d_{jk},\label{eq: popularity_effect}
\end{equation}
which implies that agents receive more utility from links with popular
(or high degree) agents.

The second example specifies $s_{ij}\left(\mathbf{d}\right)$ as

\begin{equation}
s_{ij}\left(\mathbf{d}\right)=\sum_{k}d_{ik}d_{jk},\label{eq: transitivity_effect}
\end{equation}
which implies that dyads receive more utility from linking when they
share other links in common. This is a transitivity effect.

When the utility function is of the form given in (\ref{eq: utility_function})
the marginal utility agent $i$ gets from a link with $j$ is
\[
MU_{ij}\left(\mathbf{d},\mathbf{U};\gamma_{0},\delta_{0}\right)=A_{i}+B_{j}+\gamma_{0}s_{ij}\left(\mathbf{d}\right)-U_{ij}.
\]
Pairwise stability then implies that, conditional on the realizations
of $\mathbf{A}$, $\mathbf{B}$, $\mathbf{U}$ and the value of externality
parameter $\gamma_{0}$, the observed network must satisfy, for $i=1,\ldots,N-1$
and $j=i+1,\ldots,N$
\begin{equation}
D_{ij}=\mathbf{1}\left(\tilde{A}_{i}+\tilde{A}_{j}+\gamma_{0}\tilde{s}_{ij}\left(\mathbf{d}\right)\geq\tilde{U}_{ij}\right)\label{eq: link_rule_incomplete}
\end{equation}
with $\tilde{A}_{i}=A_{i}+B_{i}$, $\tilde{s}_{ij}\left(\mathbf{d}\right)=s_{ij}\left(\mathbf{d}\right)+s_{ji}\left(\mathbf{d}\right)$
and $\tilde{U}_{ij}=U_{ij}+U_{ji}$.

Equation (\ref{eq: link_rule_incomplete}) defines a system of $\tbinom{N}{2}=\frac{1}{2}N\left(N-1\right)$
nonlinear simultaneous equations. Any solution to this system --
and there will typically be multiple ones -- constitutes a pairwise
stable (with transfers) network. To make this observation a bit more
explicit, similar to \citet{Miyauchi_JOE16}, consider the mapping
$\varphi\left(\mathbf{D}\right)\thinspace:\thinspace\mathbb{D}_{N}\rightarrow\mathbb{I}_{\tbinom{N}{2}}$:
\begin{equation}
\varphi\left(\mathbf{d}\right)\equiv\left[\begin{array}{c}
\mathbf{1}\left(\tilde{A}_{1}+\tilde{A}_{2}+\gamma_{0}\tilde{s}_{12}\left(\mathbf{d}\right)\geq U_{12}\right)\\
\mathbf{1}\left(\tilde{A}_{1}+\tilde{A}_{3}+\gamma_{0}\tilde{s}_{13}\left(\mathbf{d}\right)\geq U_{13}\right)\\
\vdots\\
\mathbf{1}\left(\tilde{A}_{N-1}+\tilde{A}_{N}+\gamma_{0}\tilde{s}_{N-1N}\left(\mathbf{d}\right)\geq U_{N-1N}\right)
\end{array}\right].\label{eq: fixed_point}
\end{equation}
Under the maintained assumption that the observed network satisfies
Definition \ref{def: Pairwise-stability}, the observed adjacency
matrix corresponds to the fixed point
\[
\mathbf{D}=\text{vech}^{-1}\left[\varphi\left(\mathbf{D}\right)\right].
\]
Here $\mathrm{vech}(\cdot)$ vectorizes the $\tbinom{N}{2}$ elements
in the lower triangle of an $N\times N$ matrix and we define its
inverse operator as creating a symmetric matrix with a zero diagonal.
In addition to the observed network there may be other $\mathbf{d}\in\mathbb{D}_{N}$
such that $\mathbf{d}=\text{vech}^{-1}\left[\varphi\left(\mathbf{d}\right)\right]$.
The fixed point representation is useful for showing equilibrium existence
as well as for characterization (e.g., using Tarski's \citeyearpar{Tarski_PJM55}
fixed point theorem).

For the two types of network formation externalities we consider,
specified in equations (\ref{eq: popularity_effect}) and (\ref{eq: transitivity_effect})
above, the $\varphi\left(\mathbf{d}\right)$ mapping is weakly increasing
in $\mathbf{d}$ for $\gamma_{0}\geq0$. This allows for the application
of Tarski's \citeyearpar{Tarski_PJM55} theorem; ensuring existence
of at least one pairwise stable equilibrium.

\section*{Test formulation}

Our goal is to assess the null hypothesis that $\gamma_{0}=0$ relative
to the alternative that $\gamma_{0}>0$. The extension of what follows
to two-sided tests is straightforward. A feature our testing problem
is the presence of a high-dimensional nuisance parameter in the form
of the $N$ degree heterogeneity terms, $\mathbf{\tilde{A}}=\left[\tilde{A}_{i}\right]$.
Since the value of these terms may range freely over the null, our
null hypothesis is a \emph{composite} one.

The composite nature of the null hypothesis raises concerns about
size control. Ideally our test will have good size properties regardless
of the particular value of $\mathbf{\tilde{A}}$. Assume, for example,
that the distribution of the $\left\{ \tilde{A}_{i}\right\} _{i=1}^{N}$
is right-skewed. In this case we will likely observe high levels of
clustering among high $\tilde{A}_{i}$ agents. Measured transitivity
in the network might be substantial even in the absence of any structural
preference for transitive relationships. We want to avoid excessive
rejection of our null hypothesis in such settings; we do so by varying
the critical value used for rejection with the magnitude of a sufficient
statistics for the $\left\{ \tilde{A}_{i}\right\} _{i=1}^{N}$.

A simple example helps to fix ideas. Under the null we have, for $i=1,\ldots,N-1$
and $j=i+1,\ldots,N$, 
\begin{equation}
\Pr\left(D_{ij}=1\right)=\frac{\exp\left(\tilde{A}_{i}+\tilde{A}_{j}\right)}{1+\exp\left(\tilde{A}_{i}+\tilde{A}_{j}\right)},\label{eq: beta_model}
\end{equation}
which corresponds to the $\beta$-model of network formation \citep[e.g.,][]{Chatterjee_et_al_AAP11}.
Assume that $\tilde{A}_{i}\in\left\{ -\infty,\frac{1}{2}\ln\left(\frac{\rho}{1-\rho}\right)\right\} $
with $\tilde{A}_{i}=-\infty$ with probability $1-\pi$. In this simple
model two ``high'' $\tilde{A}_{i}$ agents link with probability
$\rho$, while low-to-low and low-to-high links never form. Some simple
calculations give an overall density for this network of $\pi^{2}\rho$,
and a (population) transitivity index of $\rho$. For $\pi$ small
and or $\rho$ large, transitivity in this network may exceed density
substantially even though there is no structural taste for transitive
links among agents. Here the transitivity is entirely generated by
high degree agents linking with one another with greater frequency
and, only incidentally, forming triangles in the process. A simple
comparison of density and transitivity in this case is uninformative.

Motivated in part by this inferential challenge, as well as to exploit
classic results on testing in exponential families \citep[e.g.,][Chapter 4]{Lehmann_Romano_TSH05},
it will be convenient in what follows to assume that $\tilde{U}_{ij}\sim\text{Logistic}\left(0,1\right)$.
Next let $\mathbb{A}$ denoting a subset of the $N$ dimensional Euclidean
space in which $\tilde{\mathbf{A}}$ is, a priori, known to lie, and
\[
\Theta_{0}=\left\{ \left(\gamma,\tilde{\mathbf{A}}'\right)\thinspace:\thinspace\gamma=0,\tilde{\mathbf{A}}\in\mathbb{A}\right\} .
\]
Our null hypothesis is
\begin{equation}
H_{0}\thinspace:\thinspace\theta\in\Theta_{0}\label{eq: composite_null}
\end{equation}
since $\tilde{\mathbf{A}}$ may range freely over $\mathbb{A}\subset\mathbb{R}^{N}$
under the null of no externalities in link formation ($\gamma_{0}=0$).\footnote{There is an additional (implicit) nuisance parameter associated with
equilibrium selection since, under the alternative, there may be many
pairwise stable network configurations. We can ignore this complication
for our present purposes, but see \citet{Pelican_Graham_WP2019} for
additional discussion and details.} With a little manipulation we can show that, under (\ref{eq: composite_null}),
the probability of the event $\mathbf{D}=\mathbf{d}$ takes the exponential
family form
\[
P_{0}\left(\mathbf{d};\tilde{\mathbf{A}}\right)=c\left(\tilde{\mathbf{A}}\right)\exp\left(\mathbf{d}_{+}'\tilde{\mathbf{A}}\right)
\]
with $\mathbf{d}_{+}=\left(d_{1+},\ldots,d_{N+}\right)$ equal to
the degree sequence of the network.

Let $\mathbb{D}_{N,\mathbf{\mathbf{d}_{+}}}$ denote the set of all
undirected $N\times N$ adjacency matrices with degree counts also
equal to $\mathbf{d}_{+}$ and $\left|\mathbb{D}_{N,\mathbf{\mathbf{d}_{+}}}\right|$
denote the size, or cardinality, of this set. Under $H_{0}$ the conditional
likelihood of $\mathbf{D}=\mathbf{d}$ given $\mathbf{D_{+}=d_{+}}$
is
\[
P_{0}\left(\left.\mathbf{d}\right|\mathbf{D_{+}=d_{+}}\right)=\frac{1}{\left|\mathbb{D}_{N,\mathbf{\mathbf{d}_{+}}}\right|}.
\]
Under the null of no externalities \emph{all networks with identical
degree sequences are equally probable}. This insight will form the
basis of our test.

Let $T\left(\mathbf{d}\right)$ be some statistic of the adjacency
matrix $\mathbf{D}=\mathbf{d}$, say its transitivity index. We work
with a (test) critical function of the form
\[
\phi\left(\mathbf{d}\right)=\left\{ \begin{array}{cc}
1 & T\left(\mathbf{d}\right)>c_{\alpha}\left(\mathbf{d}_{+}\right)\\
g_{\alpha}\left(\mathbf{d}_{+}\right) & T\left(\mathbf{d}\right)=c_{\alpha}\left(\mathbf{d}_{+}\right)\\
0 & T\left(\mathbf{d}\right)<c_{\alpha}\left(\mathbf{d}_{+}\right)
\end{array}\right..
\]
We will reject the null if our statistic exceeds some critical value,
$c_{\alpha}\left(\mathbf{d}_{+}\right)$ and accept it -- or fail
to reject it -- if our statistic falls below this critical value.
If our statistic exactly equals the critical value, then we reject
with probability $g_{\alpha}\left(\mathbf{d}_{+}\right)$. The critical
value $c_{\alpha}\left(\mathbf{d}_{+}\right)$, as well as the probability
$g_{\alpha}\left(\mathbf{d}_{+}\right)$, are chosen to set the rejection
probability of our test under the null equal to $\alpha$ (i.e., to
control size). In order to find the appropriate values of $c_{\alpha}\left(\mathbf{d}_{+}\right)$
and $g_{\alpha}\left(\mathbf{d}_{+}\right)$ we need to know the distribution
of $T\left(\mathbf{D}\right)$ under the null.

Conceptually this distribution is straightforward to characterize;
particularly if we proceed conditional on the degree sequence observed
in the network in hand. Under the null all possible adjacency matrices
with degree sequence $\mathbf{d}_{+}$ are equally probable. The null
distribution of $T\left(\mathbf{D}\right)$ therefore equals its distribution
across all these matrices. By enumerating all the elements of $\mathbb{D}_{N,\mathbf{\mathbf{d}_{+}}}$
and calculating $T\left(\mathbf{d}\right)$ for each one, we could
directly -- and exactly -- compute this distribution. In practice
this is not (generally) computationally feasible. Even for networks
that include as few as 10 agents, the set $\mathbb{D}_{N,\mathbf{\mathbf{d}_{+}}}$
may have millions of elements (see, for example, Table 1 of \citet{Blitzstein_Diaconis_IM11}).
Below we show how to approximate the null distribution of $T\left(\mathbf{D}\right)$
by simulation, leading to a practical method of finding critical values
for testing.

If we could efficiently enumerate the elements of $\mathbb{D}_{N,\mathbf{\mathbf{d}_{+}}}$
we would find $c_{\alpha}\left(\mathbf{d}_{+}\right)$ by solving
\begin{equation}
\Pr\left(\left.T\left(\mathbf{D}\right)\geq c_{\alpha}\left(\mathbf{d}_{+}\right)\right|\mathbf{D}\in\mathbb{D}_{\mathbf{N,\mathbf{d}_{+}}}\right)=\frac{\sum_{\mathbf{D}\in\mathbb{D}_{\mathbf{N,\mathbf{d}_{+}}}}\mathbf{1}\left(T\left(\mathbf{D}\right)\geq c_{\alpha}\left(\mathbf{d}_{+}\right)\right)}{\left|\mathbb{D}_{N,\mathbf{\mathbf{d}_{+}}}\right|}=\alpha.\label{eq: critical_value_choice}
\end{equation}
If there is no $c_{\alpha}\left(\mathbf{d}_{+}\right)$ for which
(\ref{eq: critical_value_choice}) exactly holds, then we would instead
find the smallest $c_{\alpha}\left(\mathbf{d}_{+}\right)$ such that
$\Pr\left(\left.T\left(\mathbf{D}\right)>c_{\alpha}\left(\mathbf{d}_{+}\right)\right|\mathbf{D}\in\mathbb{D}_{\mathbf{N,\mathbf{d}_{+}}}\right)\geq\alpha$
and choose $g_{\alpha}\left(\mathbf{d}_{+}\right)$ to ensure correct
size.

Alternatively we might instead calculate the p-value:
\begin{equation}
\Pr\left(\left.T\left(\mathbf{D}\right)\geq T\left(\mathbf{d}_{+}\right)\right|\mathbf{D}\in\mathbb{D}_{\mathbf{N,\mathbf{d}_{+}}}\right)=\frac{\sum_{\mathbf{D}\in\mathbb{D}_{\mathbf{N,\mathbf{d}_{+}}}}\mathbf{1}\left(T\left(\mathbf{D}\right)\geq T\left(\mathbf{d}_{+}\right)\right)}{\left|\mathbb{D}_{N,\mathbf{\mathbf{d}_{+}}}\right|}.\label{eq: p-value_definition}
\end{equation}

If this probability is very low, say less than 5 percent of all networks
in $\mathbb{D}_{\mathbf{N,\mathbf{d}_{+}}}$ have a transitivity index
larger than the one observed in the network in hand, then we might
conclude that our network is ``unusual'' and, more precisely, that
it is \emph{not} a uniform random draw from $\mathbb{D}_{\mathbf{N,\mathbf{d}_{+}}}$
(our null hypothesis).

Below we show how to approximate the probabilities to the left of
the equalities in (\ref{eq: critical_value_choice}) and (\ref{eq: p-value_definition})
by simulation.

\subsection*{Similarity of the test}

In our setting, a test $\phi\left(\mathbf{D}\right)$, will have size
$\alpha$ if its null rejection probability (NRP) is less than or
equal to $\alpha$ for \emph{all }values of the nuisance parameter:
\[
\underset{\theta\in\Theta_{0}}{\sup}\thinspace\mathbb{E}_{\theta}\left[\phi\left(\mathbf{D}\right)\right]=\underset{\mathbf{\tilde{A}}\in\mathbb{A}}{\sup}\thinspace\mathbb{E}_{\theta}\left[\phi\left(\mathbf{D}\right)\right]=\alpha.
\]
Since $\tilde{\mathbf{A}}$ is high dimensional, size control is non-trivial.
This motivates, as we have done, proceeding conditionally on the degree
sequence.

Let $\mathbb{D}_{+}$ be the set of all graphical degree sequences
(see below for a discussion of ``graphic'' integer sequences). For
each $\mathbf{d_{+}}\in\mathbb{D_{+}}$ our approach is equivalent
to forming a test with the property that, for all $\theta\in\Theta_{0}$,
\begin{equation}
\mathbb{E}_{\theta}\left[\left.\phi\left(\mathbf{D}\right)\right|\mathbf{D_{+}}=\mathbf{d}_{+}\right]=\alpha.\label{eq: conditional_size}
\end{equation}
Such an approach ensures \emph{similarity} of our test since, by iterated
expectations 
\[
\mathbb{E}_{\theta}\left[\phi\left(\mathbf{D}\right)\right]=\mathbb{E}_{\theta}\left[\mathbb{E}_{\theta}\left[\left.\phi\left(\mathbf{D}\right)\right|\mathbf{\mathbf{D_{+}}}\right]\right]=\alpha
\]
for any $\theta\in\Theta_{0}$ \citep[cf.,][]{Fergusun_MS67}. By
proceeding conditionally we ensure the NRP is unaffected by the value
of the degree heterogeneity distribution $\left\{ \tilde{A}_{i}\right\} _{i=1}^{N}$.
Similar tests have proved to be attractive in other settings with
composite null hypotheses \citep[cf.,][]{Moreira_JOE09}.

\subsection*{Choosing the test statistic}

Ideally the critical function is chosen to maximize the probability
of correctly rejecting the null under particular alternatives of interest.
It turns out that, because our network formation model is incomplete
under the alternative (as we have been silent about equilibrium selection),
constructing tests with good power is non-trivial. In \citet{Pelican_Graham_WP2019}
we show how to choose the critical function, or equivalently the statistic
$T\left(\mathbf{d}\right)$, to maximize power against particular
(local) alternatives. The argument is involved, so here we confine
ourselves to a more informal development.

An common approach to choosing a test statistic, familiar from other
applications of randomization testing \citep[e.g.,][Chapter 3]{Cox_Bk06},
is to proceed heuristically. This suggests, for example, choosing
$T\left(\mathbf{D}\right)$ to be the transitivity index, or the support
measure of \citet{Jackson_et_al_AER12}, if the researcher is interested
in ``testing'' for whether agents prefer transitive relationships.

A variation on this approach, inspired by the more formal development
in \citet{Pelican_Graham_WP2019}, is to set $T\left(\mathbf{D}\right)$
equal to
\begin{equation}
T\left(\mathbf{D}\right)=\sum_{i<j}\left(D_{ij}-\hat{p}_{ij}\right)\tilde{s}_{ij}\left(\mathbf{d}\right)\label{eq: optimal_statistic}
\end{equation}
with $\hat{p}_{ij}=\frac{\exp\left(\hat{\tilde{A}}_{i}+\hat{\tilde{A}}_{j}\right)}{1+\exp\left(\hat{\tilde{A}}_{i}+\hat{\tilde{A}}_{j}\right)}$
and $\hat{\tilde{\mathbf{A}}}=\left[\hat{\tilde{A}}_{i}\right]$ the
maximum likelihood estimate (MLE) of $\tilde{\mathbf{A}}$.\footnote{\citet{Chatterjee_et_al_AAP11} present a simple fixed point algorithm
for computing this MLE (see also \citet{Graham_EM17}).} The intuition behind (\ref{eq: optimal_statistic}) is as follows:
if it is positive, this implies that high values of the externality
term, $\tilde{s}_{ij}\left(\mathbf{d}\right)$, are associated with
links that have low estimated probability under the null (such that
$D_{ij}-\hat{p}_{ij}$ is large). The conjunction of ``surprising''
links with large values of $\tilde{s}_{ij}\left(\mathbf{d}\right)$
is taken as evidence that $\gamma_{0}>0$.

Consider the transitivity example with $\tilde{s}_{ij}\left(\mathbf{d}\right)=2\sum_{k}d_{ik}d_{jk}$;
statistic (\ref{eq: optimal_statistic}), with some manipulation,
can be shown to equal
\begin{align}
T\left(\mathbf{D}\right) & =6\left[\sum_{i<j<k}D_{ij}D_{ik}D_{jk}-\sum_{i<j<k}\frac{1}{3}\left(\hat{p}_{ij}D_{ik}D_{jk}+D_{ij}\hat{p}_{ik}D_{jk}+D_{ij}D_{ik}\hat{p}_{jk}\right)\right].\label{eq: transitivity_test_statistic}
\end{align}
Statistic (\ref{eq: transitivity_test_statistic}) is a measure of
the difference between the actual number of triangles in the network
and (a particular) expected triangle count computed under the null.
To see this assume, as would be approximately true if the graph were
an Erdös-Rényi one, that $\hat{p}_{ij}=\text{\ensuremath{\hat{P}\left(\vcenter{\hbox{\includegraphics[scale=0.125]{edge.eps}}}\right)}}$
for all $i<j$. Recalling that $\hat{P}\left(\vcenter{\hbox{\includegraphics[scale=0.125]{triangle.eps}}}\right)$
and $\hat{P}\left(\vcenter{\hbox{\includegraphics[scale=0.125]{twostar.eps}}}\right)$
respectively equal the fraction of all $\tbinom{N}{3}$ triads which
are triangles and two-stars, we would have
\[
T\left(\mathbf{D}\right)=2\tbinom{N}{3}\left[3\hat{P}\left(\vcenter{\hbox{\includegraphics[scale=0.125]{triangle.eps}}}\right)-\left(3\hat{P}\left(\vcenter{\hbox{\includegraphics[scale=0.125]{triangle.eps}}}\right)+\hat{P}\left(\vcenter{\hbox{\includegraphics[scale=0.125]{twostar.eps}}}\right)\right)\hat{P}\left(\vcenter{\hbox{\includegraphics[scale=0.125]{edge.eps}}}\right)\right].
\]

The term in $\left[\right]$ equals the difference between the numerator
of the transitivity index and its denominator \emph{times} density.
For an Erdös-Rényi graph this difference should be approximately zero.
In the presence of degree heterogeneity, the second term to the right
of the equality in (\ref{eq: transitivity_test_statistic}) is a null-model-assisted
count of the expected number of triangles in $G$. A rejection therefore
occurs when many ``surprising'' triangles are present.

Before describing how to simulate the null distribution of $T\left(\mathbf{D}\right)$
we briefly recap. We start by specifying a sharp null hypothesis.
Consider the network in hand with adjacency matrix $\mathbf{D}=\mathbf{d}$
and corresponding degree sequence $\mathbf{D}_{+}=\mathbf{d}_{+}$.
Our null hypothesis is that the observed network coincides with a
uniform random draw from $\mathbb{D}_{N,\mathbf{\mathbf{d}_{+}}}$
(i.e., the set of all networks with identical degree sequences). This
null is a consequence of the form on the transferable utility network
formation game outlined earlier. The testing procedure is to compare
a particular statistic of $\mathbf{D}=\mathbf{d}$, say $T\left(\mathbf{d}\right)$,
with its distribution across $\mathbb{D}_{N,\mathbf{\mathbf{d}_{+}}}$.
If the observed value of $T\left(\mathbf{d}\right)$ is unusually
large we take this as evidence against our null.

Although we have motivated our test as one for strategic interactions
or externalities, in actuality we are really assessing the adequacy
of a particular null model of network formation -- namely the $\beta$-model.
Our test may detect many types of violations of this model, albeit
with varying degrees of power. Consequently we need to be careful
about how we interpret a rejection in practice. At the same time,
by choosing the test statistic $T\left(\mathbf{D}\right)$ with some
care, we hope to generate good power to detect the violation of interest
-- that $\gamma_{0}>0$ -- and hence conclude that externalities
in link formation are likely present when we reject.

\section*{Simulating undirected networks with fixed degree}

This section describes an algorithm, introduced by \citet{Blitzstein_Diaconis_IM11},
for sampling uniformly from the set $\mathbb{D}_{N,\mathbf{\mathbf{d}_{+}}}$.
Our notation and exposition tracks that of \citet{Blitzstein_Diaconis_IM11},
albeit with less details. As noted previously, direct enumeration
of all the elements of $\mathbb{D}_{N,\mathbf{\mathbf{d}_{+}}}$ is
generally not feasible. We therefore require a method of sampling
from $\mathbb{D}_{N,\mathbf{\mathbf{d}_{+}}}$ \emph{uniformly} and
also, at least implicitly, estimating its size. The goal is to replace,
for example, the exact p-value (\ref{eq: p-value_definition}) with
the simulation estimate
\begin{equation}
\hat{\Pr}\left(\left.T\left(\mathbf{D}\right)\geq T\left(\mathbf{d}\right)\right|\mathbf{D}\in\mathbb{D}_{\mathbf{N,\mathbf{d}_{+}}}\right)=\frac{1}{B}\sum_{b=1}^{B}\mathbf{1}\left(T\left(\mathbf{D}_{b}\right)\geq T\left(\mathbf{d}\right)\right),\label{eq: ch3_ideal_simulation_estimate}
\end{equation}
where $\mathbf{D}_{b}$ is a uniform random draw from $\mathbb{D}_{N,\mathbf{\mathbf{d}_{+}}}$
and $B$ denotes the number of independent simulation draws selected
by the researcher.

Two complications arise. First, it is not straightforward to construct
a random draw from $\mathbb{D}_{N,\mathbf{\mathbf{d}_{+}}}$. Second,
we must draw \emph{uniformly} from this set. Fortunately the first
challenge is solvable using ideas from the discrete math literature.
Researchers in graph theory and discrete math have studied the construction
of graphs with fixed degrees and, in particular, provided conditions
for checking whether a particular degree sequence is graphical \citep*[e.g.,][]{Sierksma_Hoogeveen_JGT91}.
We say that $\mathbf{d}_{+}$ is \emph{graphical} if there is feasible
undirected network with degree sequence $\mathbf{d}_{+}$. Not all
integer sequences are graphical. The reader can verify, for example,
that there is no feasible undirected network of three agents with
degree sequence $\mathbf{d}_{+}=\left(3,2,1\right)$.

As for the second complication, although we can not easily/directly
construct a uniform random draw from $\mathbb{D}_{N,\mathbf{\mathbf{d}_{+}}}$,
we can use importance sampling \citep[e.g.,][]{Owen_MCB13} to estimate
expectations with respect to this distribution.

The basic idea and implementation is due to \citet{Blitzstein_Diaconis_IM11}.
A similar, and evidently independently derived, algorithm is presented
in \citet{DelGenio_et_al_Plos10}. While computationally faster approaches
are now available, we nevertheless present the method introduced by
\citet{Blitzstein_Diaconis_IM11} for it pedagogical value and easy
implementation. Their approach is adequate for small to medium sized
problems. Readers interested in applying the methods outlined below
to large sparse graphs might consult \citet{Rao_et_al_Sankhya96},
\citet{McDonald_et_al_SN07} or \citet{Zang_Chen_JASA13}. \citet{Pelican_Graham_WP2019}
introduce a more complicated MCMC simulation algorithm that holds
additional graph statistics constant (besides the degree sequence).
They also provide references to the fairly extensive literature on
adjacency matrix simulation.

While our presentation of the \citet{Blitzstein_Diaconis_IM11} algorithm
is motivated by a particular formal testing problem, our view is that
it is also useful for more informally finding ``unusual'' or ``interesting''
features of a given network. Are links more transitive than one would
expect in networks with similar degree sequences? Is average path
length exceptionally short? For this reason, the material presented
below may also enter a researcher's workflow during the data summarization
or exploratory analysis stage.

\subsection*{The algorithm}

A sequential network construction algorithm begins with a matrix of
zeros and sequentially adds links to it until its rows and columns
sum to the desired degree sequence. Unfortunately, unless the links
are added appropriately, it is easy to get ``stuck'' (in the sense
that at a certain point in the process it becomes impossible to reach
a graph with the desired degree and the researcher must restart the
process). The paper by \citet{Snijders_PsyMet91} provides examples
and discussion of this phenomena.

As an example consider the graphical degree sequence $\mathbf{d}_{+}=(2,2,1,1).$
If we begin with an empty graph and add an edge between agents 3 and
4, we will go from the degree sequence $(2,2,1,1)$ to a residual
one of $(2,2,0,0)$. Unfortunately $(2,2,0,0)$ is not graphical.
Adding more edges requires introducing self-loops or a double-edge,
neither of which is allowed.

Intuitively we can avoid this phenomenon by first connecting high
degree agents. \citet*{Havel_CPPM55} and \citet*{Hakimi_SIAM62}
showed that this idea works for any degree sequence
\begin{thm}
\textsc{(}\textsc{\uline{Havel-Hakimi}}\textsc{)}\label{thm: Havel_Hakimi}
Let $d_{i+}>0$, if $\mathbf{d}_{+}$ does not have at least $d_{i+}$
positive entries other than $i$ it is not graphical. Assume this
condition holds. Let $\mathbf{\tilde{d}}_{+}$ be a degree sequence
of length $N-1$ obtained\emph{ }\textup{\emph{by}}\\
\textup{\emph{\qquad{}}}\textup{{[}i{]}}\textup{\emph{\quad{}deleting
the $i^{th}$ entry of $\mathbf{d}_{+}$ and}}\\
\textup{\emph{\qquad{}}}\textup{{[}ii{]}}\textup{\emph{\quad{}}}subtracting
1 from each of the $d_{i+}$ highest elements in $\mathbf{d}_{+}$
(aside from the $i^{th}$ one).\textup{}\\
\textup{\emph{$\mathbf{d}_{+}$ is graphical if and only if $\mathbf{\tilde{d}}_{+}$
is graphical. If $\mathbf{d}_{+}$ is graphical, then it has a realization
where agent $i$ is connected to any of the $d_{i+}$ highest degree
agents (other than $i$).}}
\end{thm}
Theorem \ref{thm: Havel_Hakimi} gives a verifiable condition for
whether a degree sequence is graphical. \citet{Blitzstein_Diaconis_IM11}
extended this condition so that we can check whether a degree sequence
is graphical if one node is already connected to some other nodes.
This modified condition serves as a tool in their importance sampling
algorithm.

Theorem \ref{thm: Havel_Hakimi} is suggestive of a sequential approach
to building an undirected network with degree sequence $\mathbf{d}_{+}$.
The procedure begins with a target degree sequence $\mathbf{d}_{+}$.
It starts by choosing a link partner for the lowest degree agent (with
at least one link). It chooses a partner for this agent from among
those with higher degree. A one is then subtracted from the lowest
degree agent and her chosen partner's degrees. This procedure continues
until the\textbf{ }\emph{residual degree sequence} (the sequence of
links that remain to be chosen for each agent) is a vector of zeros.

To formally describe such an approach we require some additional notation.
Let $\left(\oplus_{i_{1},\ldots,i_{k}}\mathbf{d}_{+}\right)$ be the
vector obtained by adding a one to the $i_{1},\ldots,i_{k}$ elements
of $\mathbf{d}_{+}$:
\[
\left(\oplus_{i_{1},\ldots,i_{k}}\mathbf{d}_{+}\right)_{j}=\left\{ \begin{array}{ll}
d_{j+}+1 & \mbox{for}\,j\in\left\{ i_{1},\ldots,i_{k}\right\} \\
d_{j+} & \mbox{otherwise}
\end{array}\right.
\]
Let $\left(\ominus_{i_{1},\ldots,i_{k}}\mathbf{d}_{+}\right)$ be
the vector obtained by subtracting one from the $i_{1},\ldots,i_{k}$
elements of $\mathbf{d}_{+}$:
\[
\left(\ominus_{i_{1},\ldots,i_{k}}\mathbf{d}_{+}\right)_{j}=\left\{ \begin{array}{ll}
d_{j+}-1 & \mbox{for}\,j\in\left\{ i_{1},\ldots,i_{k}\right\} \\
d_{j+} & \mbox{otherwise}
\end{array}\right.
\]

\begin{lyxalgorithm}
\textsc{\uline{(Blitzstein and Diaconis)}}\label{algorithm: Blizstein_Diaconis}A
sequential algorithm for constructing a random graph with degree sequence
$\mathbf{d}_{+}=\left(d_{1+},\ldots,d_{N+}\right)'$ is
\end{lyxalgorithm}
\begin{enumerate}
\item Let \textbf{$\mathbf{D}$ }be an empty adjacency matrix.
\item If $\mathbf{d}_{+}=\mathbf{0}$ terminate with output $\mathbf{D}$
\item Choose the agent $i$ with minimal positive degree $d_{i+}$.
\item Construct a list of candidate partners $\mathcal{J}=\left\{ j\neq i\,:\,D_{ij}=D_{ji}=0\,\mbox{and\,}\ominus_{i,j}\mathbf{d}_{+}\,\mbox{graphical}\right\} $.
\item Pick a partner $j\in\mathcal{J}$ with probability proportional to
its degree in $\mathbf{d}_{+}$.
\item Set $D_{ij}=D_{ji}=1$ and update $\mathbf{d}_{+}$ to $\ominus_{i,j}\mathbf{d}_{+}$.
\item Repeat steps 4 to 6 until the degree of agent $i$ is zero.
\item Return to step 2.
\end{enumerate}
The input for Algorithm \ref{algorithm: Blizstein_Diaconis} is the
target degree sequence $\mathbf{d}_{+}$ and the output is an undirected
adjacency matrix $\mathbf{D}\in\mathbb{D}_{\mathbf{N,\mathbf{d}_{+}}}$
(i.e., with $\mathbf{D}'\iota=\mathbf{d}_{+}$).\footnote{Here $\iota$ denotes a conformable column vector of ones.}

\begin{figure}
\caption{\label{fig: ch3_cubic_graph_N6}Cubic graph with six agents}

\begin{centering}
\includegraphics{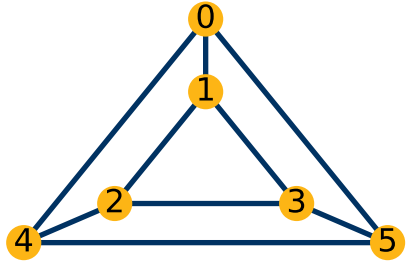}
\par\end{centering}
\textsc{\small{}\uline{Notes:}}{\small{} Prism graph (a 3-regular
graph) on six vertices.}{\small\par}

\textsc{\small{}\uline{Sources:}}{\small{} Authors' calculations.}{\small\par}
\end{figure}

Consider the 3-regular (i.e., cubic graph) depicted in Figure \ref{fig: ch3_cubic_graph_N6}.
Each agent in this graph has exactly three links such that its degree
sequence equals $\left(3,3,3,3,3,3\right)$. In turns out that there
are two non-isomorphic cubic graphs on six vertices: the prism graph,
shown in the figure, and the utility graph (or complete bipartite
graph on two sets of three vertices). We can use Algorithm \ref{algorithm: Blizstein_Diaconis}
to generate a random draw from the set of all graphs with a degree
sequence of $\mathbf{d}_{+}=\left(3,3,3,3,3,3\right)$. 

As an example of a series of residual degree sequences (updated in
Step 6 of the algorithm) associated with a random draw from $\mathbb{D}_{\mathbf{N,\mathbf{d}_{+}}}$,
for $N=6$ and $\mathbf{d}_{+}=\left(3,3,3,3,3,3\right)$, consider:
\begin{align*}
\left(\mathbf{3},3,3,3,3,3\right) & \rightarrow\left(\mathbf{2},2,3,3,3,3\right)\rightarrow\left(\mathbf{1},2,3,3,2,3\right)\rightarrow\left(0,\mathbf{2},2,3,2,3\right)\\
 & \rightarrow\left(0,\boldsymbol{1},2,3,2,2\right)\rightarrow\left(0,0,\mathbf{2},2,2,2\right)\rightarrow\left(0,0,\mathbf{1},2,1,2\right)\\
 & \rightarrow\left(0,0,0,\mathbf{2},1,1\right)\rightarrow\left(0,0,0,\mathbf{1},0,1\right)\rightarrow\left(0,0,0,0,0,0\right).
\end{align*}
Labelling agents $i=0,1,\ldots,5$ from left-to-right we can see that
the first link is added between agents 0 and 1 (the ``active'' node's
residual degree is bold-faced above). This is illustrated in Figure
\ref{fig: ch3_blitzstein_diaconis_simulation_example}, which begins
with the labelled empty graph in the upper-left-hand corner and then
sequentially adds links as we move from left-to-right and top-to-bottom.
Next a link is added between agents 0 and 4, and then between agents
0 and 2. Observe that the algorithm selected agent 0 as the lowest
degree agent in the initial step and continues to connect this vertex
with higher degree ones until all needed edges incident to it are
present.\footnote{In the event of ties for the lowest degree agent, the algorithm chooses
the one with the lowest index.}

In the 8th iteration of the algorithm an edge is added between agents
3 and 4. If, instead, an edge was added between agents 4 and 5 at
this point, the residual sequence degree sequence would have been
updated to $\left(0,0,0,2,0,0\right)$, which is not graphic. Step
4 of the algorithm prevents the addition of edges which, if added,
lead to non-graphic degree sequences. It is in this way that the algorithm
avoids getting ``stuck''. Getting stuck was a problem with earlier
approaches to binary matrix simulation, such as the method of \citet{Snijders_PsyMet91}.

\subsection*{Importance sampling}

Algorithm \ref{algorithm: Blizstein_Diaconis} produces a random draw
from $\mathbb{D}_{N,\mathbf{\mathbf{d}_{+}}}$, however, it does not
draw from this set uniformly. A key insight of \citet{Blitzstein_Diaconis_IM11}
is that one can construct importance sampling weights to correct for
non-uniformity of the draws from $\mathbb{D}_{N,\mathbf{\mathbf{d}_{+}}}$.

Let $\mathcal{\mathbb{Y}}_{N,\mathbf{d}_{+}}$ denote the set of all
possible sequences of links outputted by Algorithm \ref{algorithm: Blizstein_Diaconis}
given input $\mathbf{d}_{+}$. Let $\mathcal{D}\left(Y\right)$ be
the adjacency matrix induced by link sequence $Y$. Let $Y$ and $Y'$
be two different sequences produced by the algorithm. These sequences
are equivalent if their ``end point'' adjacency matrices coincide
(i.e., if $\mathcal{D}\left(Y\right)=\mathcal{D}\left(Y'\right)$).
We can partition $\mathcal{\mathbb{Y}}_{N,\mathbf{d}_{+}}$ into a
set of equivalence classes, the number of such classes coincides with
the number of feasible networks with degree distribution $\mathbf{d}_{+}$
(i.e., with the cardinality of $\mathbb{D}_{N,\mathbf{\mathbf{d}_{+}}}$). 

\pagebreak{}

\begin{landscape}

\begin{figure}
\caption{\label{fig: ch3_blitzstein_diaconis_simulation_example}Simulation
example}

\begin{centering}
\includegraphics[scale=0.5]{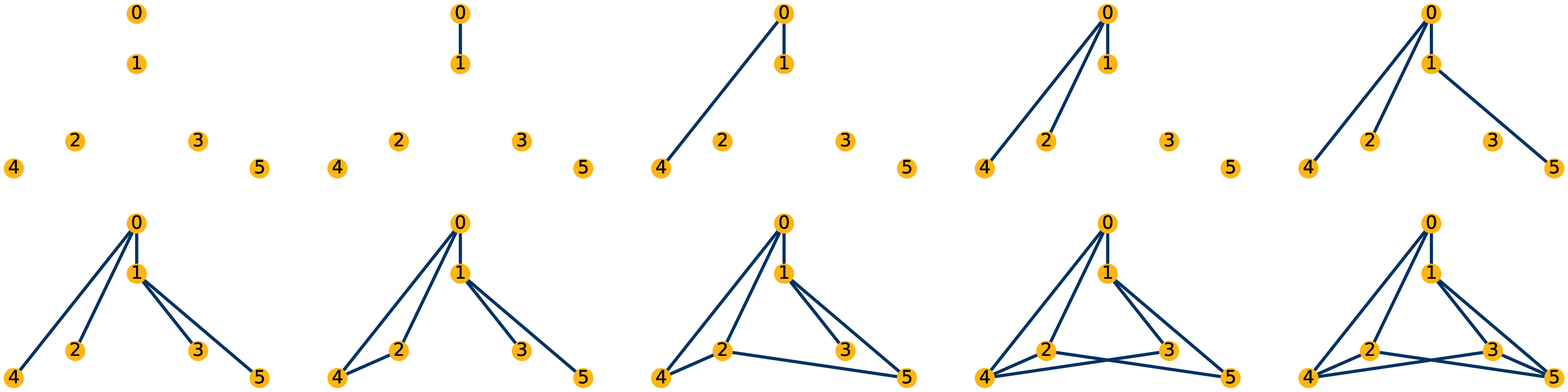}
\par\end{centering}
\textsc{\small{}\uline{Notes:}}{\small{} Illustration of the construction
of a random draw from $\mathbb{D}_{N,\mathbf{d}_{+}}$, for $N=6$
and $\mathbf{d}_{+}=\left(3,3,3,3,3,3\right)$, as generated according
to Algorithm \ref{algorithm: Blizstein_Diaconis}.}{\small\par}

\textsc{\small{}\uline{Sources:}}{\small{} Authors' calculations.}{\small\par}
\end{figure}

\pagebreak{}

\end{landscape}

Let $c\left(Y\right)$ denote the number of possible link sequences
produced by Algorithm \ref{algorithm: Blizstein_Diaconis} that produce
$Y$'s end point adjacency matrix (i.e., the number of different ways
in which Algorithm \ref{algorithm: Blizstein_Diaconis} can generate
a given adjacency matrix). Let $i_{1},i_{2},\ldots,i_{M}$ be the
sequence of agents chosen in step 3 of Algorithm \ref{algorithm: Blizstein_Diaconis}
in which $Y$ is the output. Let $a_{1},\ldots,a_{m}$ be the degree
sequences of $i_{1},\ldots,i_{M}$ at the time when each agent was
first selected in step 3, then
\begin{equation}
c\left(Y\right)=\prod_{k=1}^{M}a_{k}!\label{eq: ch3_c(Y)}
\end{equation}
To see why (\ref{eq: ch3_c(Y)}) holds consider two equivalent link
sequences $Y$ and $Y'$. Because links are added to vertices by minimal
degree (see Step 3 of Algorithm \ref{algorithm: Blizstein_Diaconis}),
the agent sequences $i_{1},i_{2},\ldots,i_{M}$ coincide for $Y$
and $Y'$. This, in turn, means that\emph{ the exact same links},
perhaps in a different order, are added at each ``stage'' of the
algorithm (i.e., when the algorithm iterates through steps 4 to 7
repeatedly for a given agent). The number of different ways to add
agent $i_{k}$'s links during such a ``stage'' is simply $a_{k}!$
and hence (\ref{eq: ch3_c(Y)}) follows.

The second component needed to construct importance weights is $\sigma\left(Y\right)$,
the probability that Algorithm \ref{algorithm: Blizstein_Diaconis}
produces link sequence $Y$. This probability is easy to compute.
Each time the algorithm chooses a link in step 5 we simply record
the probability with which it was chosen (i.e., the residual degree
of the chosen agent divided by the sum of the residual degrees of
all agents in the choice set). The product of all these probabilities
equals $\sigma\left(Y\right)$.

With $c\left(Y\right)$ and $\sigma\left(Y\right)$ defined we can
now show how to estimate expectations with respect to uniform draws
from $\mathbb{D}_{N,\mathbf{\mathbf{d}_{+}}}$. Let $T\left(\mathbf{D}\right)$
be some statistic of the adjacency matrix. Here for $\mathbf{D}$
is a draw from $\mathbb{D}_{N,\mathbf{\mathbf{d}_{+}}}$ constructed
using Algorithm \ref{algorithm: Blizstein_Diaconis}. Consider the
p-value estimation problem discussed earlier:

\begin{eqnarray*}
\mathbb{E}\left[\frac{\pi\left(\mathcal{D}\left(Y\right)\right)}{c\left(Y\right)\sigma\left(Y\right)}\mathbf{1}\left(T\left(\mathcal{D}\left(Y\right)\right)>T\left(\mathbf{d}\right)\right)\right] & = & \sum_{y\in\mathbb{Y}_{N,\mathbf{d}}}\frac{\pi\left(\mathcal{D}\left(y\right)\right)}{c\left(y\right)\sigma\left(y\right)}\mathbf{1}\left(T\left(\mathcal{D}\left(y\right)\right)>T\left(\mathbf{d}\right)\right)\sigma\left(y\right)\\
 & = & \sum_{y\in\mathbb{Y}_{N,\mathbf{d}}}\frac{\pi\left(\mathcal{D}\left(y\right)\right)}{c\left(y\right)}\mathbf{1}\left(T\left(\mathcal{D}\left(y\right)\right)>T\left(\mathbf{d}\right)\right)\\
 & = & \sum_{\mathbf{D}\in\mathbb{D}_{N,\mathbf{d_{+}}}}\sum_{\left\{ y\thinspace:\thinspace\mathcal{D}\left(y\right)=\mathbf{D}\right\} }\frac{\pi\left(\mathbf{D}\right)}{c\left(y\right)}\mathbf{1}\left(T\left(\mathbf{D}\right)>T\left(\mathbf{d}\right)\right)\\
 & = & \sum_{\mathbf{D}\in\mathbb{D}_{N,\mathbf{d_{+}}}}\pi\left(\mathbf{D}\right)\mathbf{1}\left(T\left(\mathbf{D}\right)>T\left(\mathbf{d}\right)\right)\\
 & = & \mathbb{E}_{\pi}\left[\mathbf{1}\left(T\left(\mathbf{D}\right)>T\left(\mathbf{d}\right)\right)\right].
\end{eqnarray*}
Here $\pi\left(\mathbf{D}\right)$ is the probability attached to
the adjacency matrix $\mathbf{D}\in\mathbb{D}_{N,\mathbf{d}_{+}}$
in the target distribution over $\mathbb{D}_{N,\mathbf{d}_{+}}$.
The ratio $\pi\left(\mathcal{D}\left(Y\right)\right)/c\left(Y\right)\sigma\left(Y\right)$
is called the likelihood ratio or the \emph{importance weight.}\textbf{
}We would like $\pi\left(\mathbf{D}\right)=1/\left|\mathbb{D}_{N,\mathbf{d}_{+}}\right|$
for all $\mathbf{D}\in\mathbb{D}_{N,\mathbf{d}_{+}}$.

Observe that $\mathbb{E}_{\pi}\left[T\left(\mathcal{D}\left(Y\right)\right)\right]=\mathbb{E}\left[\frac{\pi\left(\mathcal{D}\left(Y\right)\right)}{c\left(Y\right)\sigma\left(Y\right)}T\left(\mathcal{D}\left(Y\right)\right)\right]$;
setting $\pi\left(\mathcal{D}\left(Y\right)\right)=\left|\mathbb{D}_{N,\mathbf{d}_{+}}\right|^{-1}$
and $T\left(\mathcal{D}\left(Y\right)\right)=1$ to the constant statistic,
then suggests an estimate of $\left|\mathbb{D}_{N,\mathbf{d}_{+}}\right|$
equal to
\begin{equation}
\hat{\left|\mathbb{D}_{N,\mathbf{d}_{+}}\right|}=\left[\frac{1}{B}\sum_{b=1}^{B}\frac{1}{c\left(Y_{b}\right)\sigma\left(Y_{b}\right)}\right],\label{eq: estimate_of_size_of_D_N_d+}
\end{equation}
and hence a p-value estimate of
\begin{equation}
\hat{\rho}_{T\left(\mathbf{G}\right)}=\left[\frac{1}{B}\sum_{b=1}^{B}\frac{1}{c\left(Y_{b}\right)\sigma\left(Y_{b}\right)}\right]^{-1}\times\left[\frac{1}{B}\sum_{b=1}^{B}\frac{1}{c\left(Y_{b}\right)\sigma\left(Y_{b}\right)}\mathbf{1}\left(T\left(\mathbf{D}_{b}\right)>T\left(\mathbf{d}\right)\right)\right].\label{eq: estimate_of_average_of_T(D)}
\end{equation}
An attractive feature of (\ref{eq: estimate_of_average_of_T(D)})
is that the importance weights need only be estimated up to a constant.
This feature is useful when dealing with numerical overflow issues
that can arise when $\left|\mathbb{D}_{N,\mathbf{d}_{+}}\right|$
is too large to estimate.

Algorithm (\ref{algorithm: Blizstein_Diaconis}) is appropriate for
simulating undirected networks. Recently \citet{Kim_et_al_NJP12}
propose a method for simulating from directed networks with both fixed
indegree and outdegree sequences. Their methods is based on an extension
of Havel-Hakimi type results to digraphs due to \citet{Erdos_et_al_EJC10}.
\citet{Pelican_Graham_WP2019} introduce an MCMC algorithm for simulating
digraphs satisfying various side constraints.

\section*{Illustration using the Nyakatoke network}

\begin{figure}
\caption{\label{fig: Nyakatoke-Degree-Distribution}Nyakatoke Degree Distribution}

\begin{centering}
\includegraphics{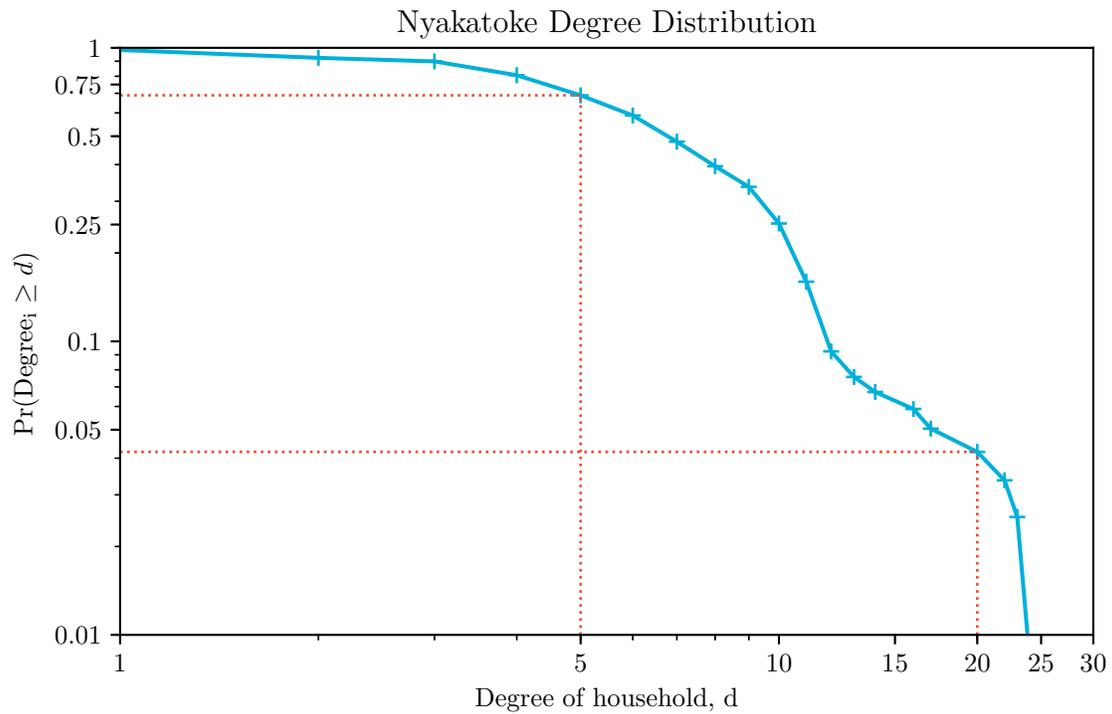}
\par\end{centering}
\textsc{\small{}\uline{Notes:}}{\small{} This figure plots the
probability (vertical axis) that a random household in Nyakatoke has
strictly more risk sharing links than listed on the horizontal axis.}{\small\par}

\textsc{\small{}\uline{Source:}}{\small{} \citet{deWeerdt_IAP04}
and authors' calculations. Raw data available at \url{https://www.uantwerpen.be/en/staff/joachim-deweerdt/}
(Accessed January 2017).}{\small\par}
\end{figure}

The transitivity index for the Nyakatoke network, at 0.1884, is almost
three times its associated network density of 0.0698. Is this excess
transitivity simply a product of degree heterogeneity alone? To assess
this we used Algorithm \ref{algorithm: Blizstein_Diaconis} to take
5,000 draws from the set of adjacency matrices with $N=119$ and degree
sequences coinciding with the one observed in Nyakatoke (for reference
the Nyakatoke degree distribution is plotted in Figure \ref{fig: Nyakatoke-Degree-Distribution}).

\begin{figure}
\caption{\label{fig: importance_sampling_transitivity}Reference distribution
of transitivity index for Nyakatoke network}

\begin{centering}
\includegraphics[scale=0.65]{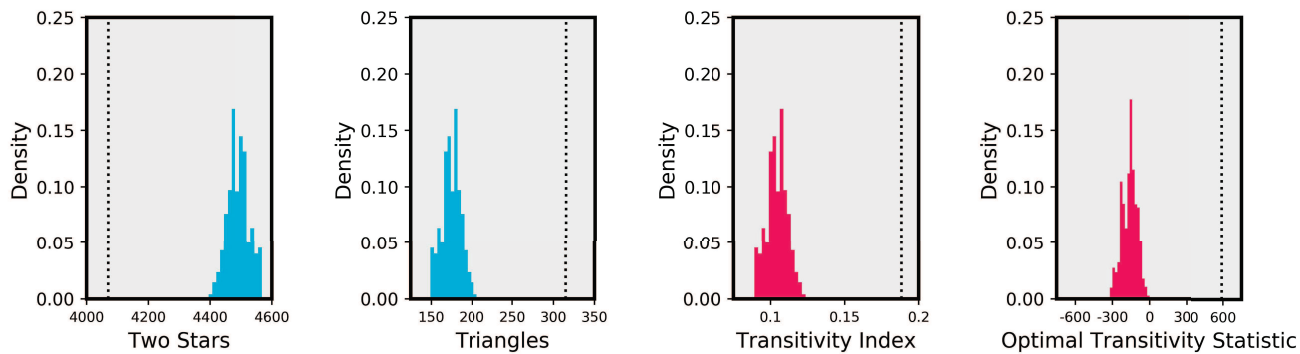}
\par\end{centering}
\textsc{\small{}\uline{Notes:}}{\small{} Histogram of two star
counts, triangle counts and transitivity index values across 5,000
draws from $\mathbb{D}_{N,\mathbf{d}_{+}}$ ($N=119$ and $\mathbf{d}_{+}$coinciding
with the one observed in Nyakatoke). The final figure plots the distribution
of the ``optimal'' transitivity statistic given in Equation (\ref{eq: optimal_statistic}).}{\small\par}

\textsc{\small{}\uline{Sources:}}{\small{} \citet{deWeerdt_IAP04}
and authors' calculations. Raw data available at \url{https://www.uantwerpen.be/en/staff/joachim-deweerdt/}
(Accessed January 2017).}{\small\par}
\end{figure}

Figure \ref{fig: importance_sampling_transitivity} displays estimates
of the distribution of two star and triangle counts, as well as the
transitivity index (and ``optimal'' transitivity statistic), with
respect to the distribution of uniform draws from $\mathbb{D}_{N,\mathbf{d}_{+}}$
($N=119$ and $\mathbf{d}_{+}$ coinciding with the one observed in
Nyakatoke). Measured transitivity is Nyakatoke is extreme relative
to this reference distribution. This suggests that clustering of links
is, in fact, a special feature of the Nyakatoke network. It is also
interesting to note that the distribution of transitivity in this
reference distribution is well to the right of 0.0698 (the density
of all graphs in the reference distribution). The skewed degree distribution
in Nyakatoke forces a certain amount of transitivity, since high degree
nodes are more likely to link with one another. This highlights the
value of a test which proceeds conditionally on the degree sequence.

\begin{figure}
\caption{\label{fig: importance_sampling_diameter}Reference distribution of
network diameter and average distance for Nyakatoke network}

\begin{centering}
\includegraphics{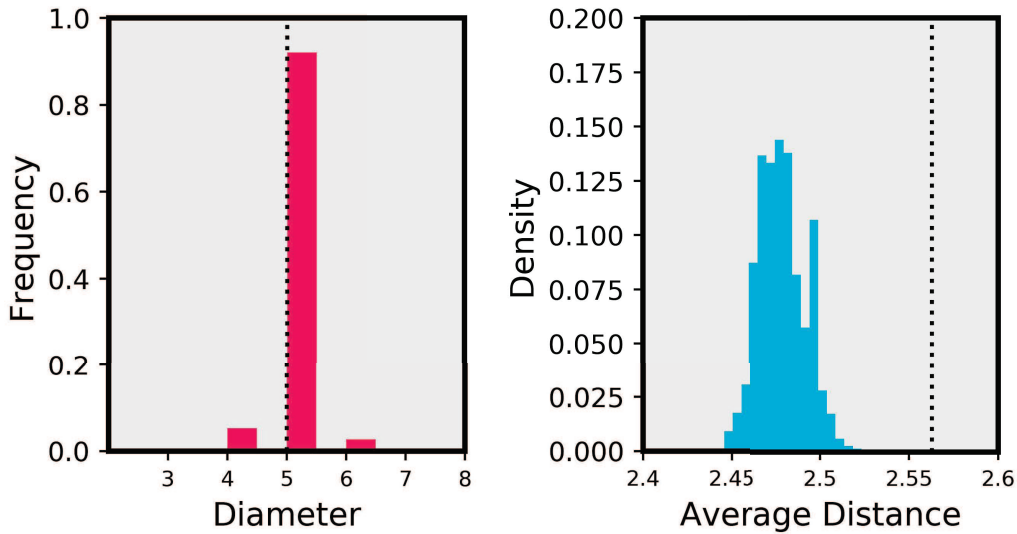}
\par\end{centering}
\textsc{\small{}\uline{Notes:}}{\small{} Histogram of values network
diameter and average distance across 5,000 draws from $\mathbb{D}_{N,\mathbf{d}_{+}}$
($N=119$ and $\mathbf{d}_{+}$coinciding with the one observed in
Nyakatoke).}{\small\par}

\textsc{\small{}\uline{Sources:}}{\small{} \citet{deWeerdt_IAP04}
and authors' calculations. Raw data available at \url{https://www.uantwerpen.be/en/staff/joachim-deweerdt/}
(Accessed January 2017).}{\small\par}
\end{figure}

Figure \ref{fig: importance_sampling_diameter} displays estimates
of the distribution of network diameter and average distance. Nyakatoke's
diameter is not atypical across networks with the same degree sequence.
However, average distance is significantly longer in Nyakatoke. One
interpretation of this fact is that the Nyakatoke includes a distinct
periphery of poorly connected/insured households.

\bibliographystyle{apalike2}
\bibliography{../Reference_BibTex/Networks_References}

\begin{thebibliography}{}

\bibitem[Blitzstein \& Diaconis, 2011]{Blitzstein_Diaconis_IM11}
Blitzstein, J. \& Diaconis, P. (2011).
\newblock A sequential importance sampling algorithm for generating random
  graphs with prescribed degrees.
\newblock {\em Internet Mathematics}, 6(4), 489 -- 522.

\bibitem[Bloch \& Jackson, 2006]{Bloch_Jackson_IJGT06}
Bloch, F. \& Jackson, M.~O. (2006).
\newblock Definitions of equilibrium in network formation games.
\newblock {\em International Journal of Game Theory}, 34(3), 305 -- 318.

\bibitem[Bloch \& Jackson, 2007]{Bloch_Jackson_JET07}
Bloch, F. \& Jackson, M.~O. (2007).
\newblock The formation of networks with transfers among players.
\newblock {\em Journal of Economic Theory}, 113(1), 83 -- 110.

\bibitem[Chatterjee et~al., 2011]{Chatterjee_et_al_AAP11}
Chatterjee, S., Diaconis, P., \& Sly, A. (2011).
\newblock Random graphs with a given degree sequence.
\newblock {\em Annals of Applied Probability}, 21(4), 1400 -- 1435.

\bibitem[Comola \& Fafchamps, 2014]{Comola_Fafchamps_EJ14}
Comola, M. \& Fafchamps, M. (2014).
\newblock Testing unilateral and bilateral link formation.
\newblock {\em Economic Journal}, 124(579), 954 -- 976.

\bibitem[Cox, 2006]{Cox_Bk06}
Cox, D.~R. (2006).
\newblock {\em Principles of Statistical Inference}.
\newblock Cambridge: Cambridge University Press.

\bibitem[De~Weerdt, 2004]{deWeerdt_IAP04}
De~Weerdt, J. (2004).
\newblock {\em Insurance Against Poverty}, chapter Risk-sharing and endogenous
  network formation, (pp.\ 197 -- 216).
\newblock Oxford University Press: Oxford.

\bibitem[Del~Genio et~al., 2010]{DelGenio_et_al_Plos10}
Del~Genio, C.~I., Kim, H., Toroczkai, Z., \& Bassler, K. (2010).
\newblock Efficient and exact sampling of simple graphs with given arbitrary
  degree sequence.
\newblock {\em Plos One}, 5(4), e100012.

\bibitem[Erd\"{o}s et~al., 2010]{Erdos_et_al_EJC10}
Erd\"{o}s, P.~L., Mikos, I., \& Toroczkai, Z. (2010).
\newblock A simple havel-hakimi type algorithm to realize graphical degree
  sequences of directed graphs.
\newblock {\em Electronic Journal of Combinatorics}, 17(1), R66.

\bibitem[Ferguson, 1967]{Fergusun_MS67}
Ferguson, T.~S. (1967).
\newblock {\em Mathematical Statistics: A Decision Theoretic Approach}.
\newblock New York: Academic Press.

\bibitem[Fisher, 1935]{Fisher_DesignBook35}
Fisher, R.~A. (1935).
\newblock {\em The Design of Experiments}.
\newblock Edinburgh: Oliver and Boyd.

\bibitem[Graham, 2016]{Graham_NBER16}
Graham, B.~S. (2016).
\newblock {\em Homophily and transitivity in dynamic network formation}.
\newblock NBER Working Paper 22186, National Bureau of Economic Research.

\bibitem[Graham, 2017]{Graham_EM17}
Graham, B.~S. (2017).
\newblock An econometric model of network formation with degree heterogeneity.
\newblock {\em Econometrica}, 85(4), 1033 -- 1063.

\bibitem[Hakimi, 1962]{Hakimi_SIAM62}
Hakimi, S.~L. (1962).
\newblock On realizability of a set of integers as degrees of the vertices of a
  linear graph. i.
\newblock {\em Journal of the Society for Industrial and Applied Mathematics},
  10(3), 496 -- 506.

\bibitem[Havel, 1955]{Havel_CPPM55}
Havel, V.~J. (1955).
\newblock A remark on the existence of finite graph.
\newblock {\em {\v C}asopis Pro P{\v e}stov{\'a}n{\'\i} Matematiky}, 80, 477 --
  480.

\bibitem[Jackson, 2008]{Jackson_NetBook08}
Jackson, M.~O. (2008).
\newblock {\em Social and Economic Networks}.
\newblock Princeton: Princeton University Press.

\bibitem[Jackson et~al., 2012]{Jackson_et_al_AER12}
Jackson, M.~O., Rodriguez-Barraquer, T., \& Tan, X. (2012).
\newblock Social capital and social quilts: network patterns of favor exchange.
\newblock {\em American Economic Review}, 102(5), 1857--1897.

\bibitem[Kim et~al., 2012]{Kim_et_al_NJP12}
Kim, H., Del~Genio, C.~I., Bassler, K.~E., \& Toroczkai, Z. (2012).
\newblock Constructing and sampling directed graphs with given degree
  sequences.
\newblock {\em New Journal of Physics}, 14, 023012.

\bibitem[Lehmann \& Romano, 2005]{Lehmann_Romano_TSH05}
Lehmann, E.~L. \& Romano, J.~P. (2005).
\newblock {\em Testing Statistical Hypotheses}.
\newblock New York: Springer, 3rd edition.

\bibitem[McDonald et~al., 2007]{McDonald_et_al_SN07}
McDonald, J.~W., Smith, P. W.~F., \& Forster, J.~J. (2007).
\newblock Markov chain monte carlo exact inference for social networks.
\newblock {\em Social Networks}, 29(1), 127 -- 136.

\bibitem[McPherson et~al., 2001]{McPherson_et_al_ARS01}
McPherson, M., Smith-Lovin, L., \& Cook, J.~M. (2001).
\newblock Birds of a feather: homophily in social networks.
\newblock {\em Annual Review of Sociology}, 27(1), 415 -- 444.

\bibitem[Milo et~al., 2002]{Milo_el_al_Sci02}
Milo, R., Shen-Orr, S., Itzkovitz, S., Kashtan, N., Chklovskii, D., \& Alon, U.
  (2002).
\newblock Network motifs: simple building blocks of complex networks.
\newblock {\em Science}, 298(5594), 824 -- 827.

\bibitem[Miyauchi, 2016]{Miyauchi_JOE16}
Miyauchi, Y. (2016).
\newblock Structural estimation of a pairwise stable network with nonnegative
  externality.
\newblock {\em Journal of Econometrics}, 195(2), 224 -- 235.

\bibitem[Moreira, 2009]{Moreira_JOE09}
Moreira, M.~J. (2009).
\newblock Tests with correct size when instruments can be arbitrarily weak.
\newblock {\em Journal of Econometrics, 152(2), 131-140}, 152(2), 131 -- 140.

\bibitem[Owen, 2013]{Owen_MCB13}
Owen, A.~B. (2013).
\newblock Monte carlo theory, methods and examples.
\newblock Book Manuscript.

\bibitem[Pelican \& Graham, 2019]{Pelican_Graham_WP2019}
Pelican, A. \& Graham, B.~S. (2019).
\newblock {\em Testing for strategic interaction in social and economic network
  formation}.
\newblock Technical report, University of California - Berkeley.

\bibitem[Rao et~al., 1996]{Rao_et_al_Sankhya96}
Rao, A.~R., Jana, R., \& Bandyopadhyay, S. (1996).
\newblock A markov chain monte carlo method for generating random
  (0,1)-matrices with given marginals.
\newblock {\em Sankhya}, 58(2), 225 -- 242.

\bibitem[Sierksma \& Hoogeveen, 1991]{Sierksma_Hoogeveen_JGT91}
Sierksma, G. \& Hoogeveen, H. (1991).
\newblock Seven criteria for integer sequences being graphic.
\newblock {\em Journal of Graph Theory}, 15(2), 223 -- 231.

\bibitem[Snijders, 1991]{Snijders_PsyMet91}
Snijders, T. A.~B. (1991).
\newblock Enumeration and simulation methods for 0-1 matrices with given
  marginals.
\newblock {\em Psychometrika}, 56(3), 397 -- 417.

\bibitem[Tarski, 1955]{Tarski_PJM55}
Tarski, A. (1955).
\newblock A lattice-theoretical fixpoint theorem and its applications.
\newblock {\em Pacific Journal of Mathematics}, 5(2), 285 -- 309.

\bibitem[Zhang \& Chen, 2013]{Zang_Chen_JASA13}
Zhang, J. \& Chen, Y. (2013).
\newblock Sampling for conditional inference on network data,.
\newblock {\em Journal of the American Statistical Association}, 108(504), 1295
  -- 1307.

\end{thebibliography}

\end{document}